\documentclass{emulateapj}
\usepackage{color}
\usepackage{threeparttable}
\usepackage{ulem}

\newcommand\ci{[C~{\sc i}]~}

\shorttitle{Rotating starburst cores}
\shortauthors{Tadaki et al.}

\begin{document}


\title{Rotating starburst cores in massive galaxies at $z=2.5$}


\author{Ken-ichi Tadaki\altaffilmark{1,2},
Tadayuki Kodama\altaffilmark{2,3}, 
Erica J. Nelson\altaffilmark{1}, 
Sirio Belli\altaffilmark{1}, 
Natascha M. F{\"o}rster Schreiber\altaffilmark{1}, 
Reinhard Genzel\altaffilmark{1,4,5},
Masao Hayashi\altaffilmark{2}, 
Rodrigo Herrera-Camus\altaffilmark{1}, 
Yusei Koyama\altaffilmark{6}, 
Philipp Lang\altaffilmark{7}, 
Dieter Lutz\altaffilmark{1}, 
Rhythm Shimakawa\altaffilmark{3}, 
Linda J. Tacconi\altaffilmark{1}, 
Hannah $\ddot{\mathrm{U}}$bler\altaffilmark{1},
Emily Wisnioski\altaffilmark{1},
Stijn Wuyts\altaffilmark{8},
Bunyo Hatsukade\altaffilmark{9}, 
Magdalena Lippa\altaffilmark{1}, 
Kouichiro Nakanishi\altaffilmark{2,3}, 
Soh Ikarashi\altaffilmark{10}, 
Kotaro Kohno\altaffilmark{9,11}, 
Tomoko L. Suzuki\altaffilmark{3}, 
Yoichi Tamura\altaffilmark{9}, and
Ichi Tanaka\altaffilmark{6}
}


\affil{\altaffilmark{1} Max-Planck-Institut f{\"u}r extraterrestrische Physik, Giessenbachstrasse, D-85748 Garching, Germany; tadaki@mpe.mpg.de}
\affil{\altaffilmark{2} National Astronomical Observatory of Japan, 2-21-1 Osawa, Mitaka, Tokyo 181-8588, Japan}
\affil{\altaffilmark{3} Department of Astronomical Science, SOKENDAI (The Graduate University for Advanced Studies), Mitaka, Tokyo 181-8588, Japan}
\affil{\altaffilmark{4} Department of Physics, Le Conte Hall, University of California, Berkeley, CA 94720, USA}
\affil{\altaffilmark{5} Department of Astronomy, Hearst Field Annex, University of California, Berkeley, CA 94720, USA}
\affil{\altaffilmark{6} Subaru Telescope, National Astronomical Observatory of Japan, 650 North A'ohoku Place, Hilo, HI 96720, USA}
\affil{\altaffilmark{7} Max-Planck-Institut f{\"u}r Astronomie, K{\"o}nigstuhl 17, D-69117 Heidelberg, Germany}
\affil{\altaffilmark{8} Department of Physics, University of Bath, Claverton Down, Bath, BA2 7AY, UK}
\affil{\altaffilmark{9} Institute of Astronomy, The University of Tokyo, 2-21-1 Osawa, Mitaka, Tokyo 181-0015, Japan}
\affil{\altaffilmark{10} Kapteyn Astronomical Institute, University of Groningen, P.O. Box 800, 9700AV Groningen, The Netherlands}
\affil{\altaffilmark{11} Research Center for the Early Universe, The University of Tokyo, 7-3-1 Hongo, Bunkyo, Tokyo 113-0033, Japan}


\begin{abstract}
We present spatially resolved ALMA observations of the CO $J=3-2$ emission line in two massive galaxies at $z=2.5$ on the star-forming main sequence.
Both galaxies have compact dusty star-forming cores with effective radii of $R_\mathrm{e}=1.3\pm0.1$ kpc and $R_\mathrm{e}=1.2\pm0.1$ kpc in the 870 $\mu$m continuum emission.
The spatial extent of star-forming molecular gas is also compact with $R_\mathrm{e}=1.9\pm0.4$ kpc and $R_\mathrm{e}=2.3\pm0.4$ kpc, but more extended than the dust emission.
Interpreting the observed position-velocity diagrams with dynamical models, we find the starburst cores to be rotation-dominated with the ratio of the maximum rotation velocity to the local velocity dispersion of $v_\mathrm{max}/\sigma_0=7.0^{+2.5}_{-2.8}$ ($v_\mathrm{max}=386^{+36}_{-32}$ km s$^{-1}$) and $v_\mathrm{max}/\sigma_0=4.1^{+1.7}_{-1.5}$ ($v_\mathrm{max}=391^{+54}_{-41}$ km s$^{-1}$).
Given that the descendants of these massive galaxies in the local universe are likely ellipticals with $v/\sigma$ nearly an order of magnitude lower,
the rapidly rotating galaxies would lose significant net angular momentum in the intervening time.
The comparisons among dynamical, stellar, gas, and dust mass suggest that the starburst CO-to-H$_2$ conversion factor of $\alpha_\mathrm{CO}=0.8~M_\odot$ (K km s$^{-1}$pc$^{-2}$)$^{-1}$ is appropriate in the spatially resolved cores.
The dense cores are likely to be formed in extreme environments similar to the central regions of local ultraluminous infrared galaxies.
Our work also demonstrates that a combination of medium-resolution CO and high-resolution dust continuum observations is a powerful tool for characterizing the dynamical state of molecular gas in distant galaxies.
\end{abstract}


\keywords{galaxies: evolution --- galaxies: high-redshift --- galaxies: ISM}

\section{Introduction}
\label{sec;intro}

Massive quiescent galaxies often have dense cores \citep{2013ApJ...776...63F,2014ApJ...791...45V} while the morphology of star-forming galaxies is typically dominated by exponential disks rather than central bulges \citep[e.g.,][]{2011ApJ...742...96W}.
Massive star-forming galaxies are expected to transform their morphology from disk-dominated to bulge-dominated.
Understanding the formation history of the bulge component is a critical step toward revealing the origin of the Hubble sequence.
At the peak epoch of galaxy formation ($z\sim2$), the most massive, $\log(M_*/M_\odot)>11$, star-forming galaxies still have extended disks, but are rapidly building up their central cores through dusty, compact starbursts \citep{2017ApJ...834..135T, 2016ApJ...827L..32B}.
Bulge formation in a short period of $<$1 Gyr at $z\sim2$ is also corroborated by observations of old stellar populations and enhanced [$\alpha$/Fe] ratios in massive quiescent galaxies at $z\sim1$ \citep[e.g.,][]{2015ApJ...799..206B, 2015ApJ...808..161O}.
All these findings suggest that central cores of massive galaxies have a different formation history than outer disks. 
The next step is to characterize the kinematics of these dense cores in the process of 
formation, which will shed light on their formation mechanisms and subsequent evolution. 

At high-redshift, the kinematics of dusty star-forming cores in massive galaxies 
are difficult to study. 
While H$\alpha$ studies with near-infrared spectrographs have made significant progress in understanding the kinematics of core formation (e.g., \citealt{2014Natur.513..394N, 2014ApJ...795..145B}; E. Wisnioski et al. 2017, in preparation), the H$\alpha$ line is not an ideal tool for investigating the kinematics of forming cores because of dust attenuation. 
Multi-wavelength high-resolution imaging and emission line maps reveal that the central regions in massive high-redshift galaxies are often strongly attenuated by dust \citep{2017arXiv170400733T, 2016ApJ...817L...9N}.
CO line observations provide a more robust means of obtaining kinematic information for dusty objects, as well as the molecular gas properties \citep[e.g.,][]{2008ApJ...680..246T, 2013ApJ...772..137I}.

In local ultraluminous infrared galaxies (ULIRGs), molecular gas is concentrated into rotating nuclear disks or rings \citep[e.g.,][]{1998ApJ...507..615D}.
Moreover, the physical condition of the gas is totally different from that in normal star-forming galaxies.
In normal star-forming regions, CO emission mainly comes from an ensemble of self-gravitating molecular clouds.
Although the CO line is typically optically thick in each virialized molecular cloud, it is possible to count the number of clouds and estimate the total molecular gas mass.
We use the CO-to-H$_2$ conversion factor of $\alpha_\mathrm{CO}=M_\mathrm{gas}/L'_\mathrm{CO}=4.36~M_\odot$ (K km s$^{-1}$pc$^{-2}$)$^{-1}$ including a correction for Helium since it is calibrated by virial mass measurements, optically thin dust emission and $\gamma$-ray observations in the Milky-Way disk (see review in \citealt{2013ARA&A..51..207B}).

\begin{figure*}[t]
\begin{center}
\includegraphics[scale=1.05]{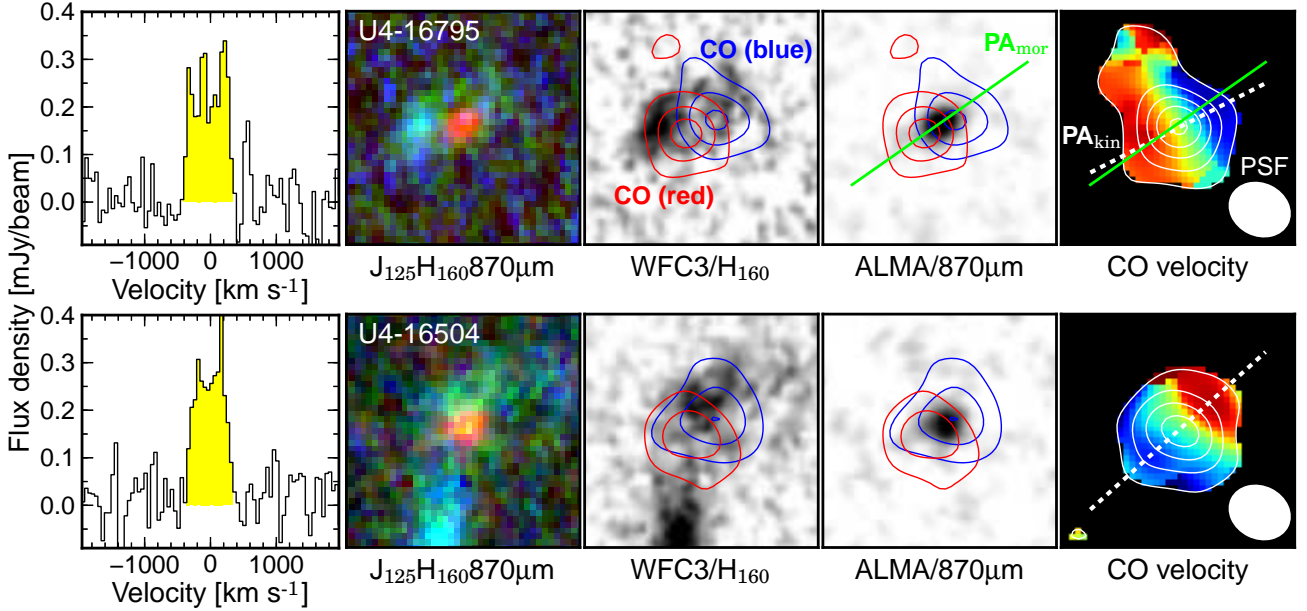}
\caption{
Two massive galaxies at $z=2.5$ detected in the CO $J=3-2$ line.
From left to right: spatially-averaged CO spectra; three-color images with HST/F125W-, F160W-, and ALMA/870 $\mu$m-band (2\farcs5$\times$2\farcs5), monochromatic images at F160W-, and 870 $\mu$m-band with blue and red contours displaying blue- and red-shifted CO components with a velocity width of 150 km s$^{-1}$; CO velocity field with white contours indicating the velocity-integrated CO fluxes.
The contours are plotted every $3\sigma$.
Green and white dashed lines indicate the morphological major axis of 870 $\mu$m continuum emission and the kinematic major axis of CO, respectively.
}
\label{fig;spectra}
\end{center}
\end{figure*}

A CO-based gas mass with the Galactic conversion factor, however, often equals or exceeds a dynamical mass in local ULIRGs and SMGs, which could imply a smaller conversion factor \citep[e.g.,][]{1998ApJ...507..615D, 2008ApJ...680..246T}.
This variation of $\alpha_\mathrm{CO}$ could be caused by a high star formation rate (SFR) surface density in extreme environments.
The intense UV radiation heats the nearby dust and the gas temperature increases through efficient energy exchange with hot dust \citep{2012MNRAS.421.3127N}.
Then, the CO surface brightness increases more rapidly than the gas mass surface density.

In this letter, we report results from CO $J = 3-2$ observations of two massive galaxies at $z=2.5$ using Atacama Large Millimeter/submillimeter Array (ALMA) to study the spatial distribution and the kinematics of molecular gas in the starburst cores.
We assume a Chabrier initial mass function (IMF; \citealt{2003PASP..115..763C}) and adopt cosmological parameters of $H_0$ =70 km s$^{-1}$ Mpc$^{-1}$, $\Omega_{\rm M}$=0.3, and $\Omega_\Lambda$ =0.7.

\section{Observations}

We focus on the most massive star-forming galaxies with $\log(M_*/M_\odot)>11$ since this mass range is important for formation of dense cores \citep[e.g.,][]{2017ApJ...834..135T}.
We select two galaxies at $z=2.53$ (U4-16795 and U4-16504) from Subaru narrow-band imaging in the SXDF field \citep{2013ApJ...778..114T}.
The narrow-band based redshift has uncertainties of $\Delta z\pm0.02$.
The two galaxies are located within the primary beam of ALMA Band-3 receivers (a beam width at half power of $\sim1$\arcmin) as the projected separation is 9\farcs7.
Both galaxies have a compact dusty star-forming core, which is probed by 870 $\mu$m dust continuum emission \citep{2017ApJ...834..135T}.
We compute the stellar mass using the {\tt FAST} spectral energy distribution fitting code \citep{2009ApJ...700..221K} and the 3D-HST multi-wavelength photometric catalog \citep{2014ApJS..214...24S} using the stellar population synthesis models of \citet{2003MNRAS.344.1000B}, exponentially declining star formation histories, and dust attenuation law of \citet{2000ApJ...533..682C}.
The total stellar mass is $\log(M_*/M_\odot)=11.26\pm0.15$ for U4-16795 and $\log(M_*/M_\odot)=11.25\pm0.15$ for U4-16504.
In deep HAWK-I/$K_s$-band maps \citep{2014A&A...570A..11F}, 81\% and 71\% of the total fluxes come from the central 1\farcs5 aperture region for U4-16795 and U4-16504, respectively.
We take into account these factors when comparing the stellar mass with other masses (Section \ref{sec;gasmass}).

U4-16795 is detected in a deep $Herschel$-PACS 160 $\mu$m map from archival data (see \citealt{2011A&A...532A..90L} for the methodology) and U4-16504 is detected in a deep $Spitzer$-MIPS 24 $\mu$m map (PI: J. Dunlop).
Following the recipes of \citet{2011ApJ...738..106W}, we derive SFRs of $\log$(SFR/$M_\odot$yr$^{-1})=2.62\pm0.1$ for U4-16795 and $\log$(SFR/$M_\odot$yr$^{-1})=2.37\pm0.25$ for U4-16504 from a combination of the rest-frame 2800 \AA~and infrared luminosities.
The targets are located on the massive end of the star-forming main sequence at $z\sim2$ \citep[e.g.,][]{2014ApJS..214...15S}.

We observe the CO $J=3-2$ emission line ($\nu_\mathrm{rest}=345.796$ GHz) of the two massive galaxies with ALMA Band-3 receivers covering the frequency range of 95--99 and 107--111 GHz.
The calibration is processed through the Common Astronomy Software Application package (CASA; \citealt{2007ASPC..376..127M}).
We use the {\tt tclean} task with natural weighting to make a channel map with a velocity width of 50 km s$^{-1}$ and dirty continuum maps excluding the frequency range of the CO line.
The synthesized beamsize is 0\farcs66$\times$0\farcs55.
The rms levels are 147 $\mu$Jy beam$^{-1}$ in the channel map and 8.1 $\mu$Jy beam$^{-1}$ in the continuum map. 

\begin{table*}
\begin{center}
\begin{threeparttable}
\caption{Galaxy properties for two massive star-forming galaxies at $z=2.5$.}\label{tab;mass}
\begin{tabular}{lccccccccc}
\hline
ID & $z_\mathrm{CO}$ & $S_\mathrm{CO}dv$\tablenotemark{a} & $\log L'_\mathrm{CO(3-2)}$ & $\log M_\mathrm{dyn}$ & $\log M_\mathrm{*}$\tablenotemark{b}  & $\log M_\mathrm{gas,CO}$\tablenotemark{b} & $\log M_\mathrm{gas,CO}$\tablenotemark{b} & $\log M_\mathrm{gas,3mm}$\tablenotemark{c} \\
& & (Jy km s$^{-1}$) & (K km s$^{-1}$pc$^{-2}$) & ($M_\odot$) &  ($M_\odot$) &  ($M_\odot$) &  ($M_\odot$)  &  ($M_\odot$) \\
& &  & & &  &  [$\alpha_\mathrm{CO}=4.36$]  & [$\alpha_\mathrm{CO}=0.8$] & [$\delta_\mathrm{gdr}=120$] \\
\hline
U4-16795 & 2.5236 & 0.72$\pm$0.06 & 10.37$\pm0.04$ & $11.10^{+0.13}_{-0.12}$ & 11.17$\pm0.15$ & 11.01$\pm0.04$ & 10.28$\pm0.04$ & $ 10.38-10.96$\\
U4-16504 & 2.5267 & 0.72$\pm$0.05 & 10.37$\pm0.03$ & $11.19^{+0.17}_{-0.15}$ & 11.10$\pm0.15$ & 11.01$\pm0.03$ & 10.27$\pm0.03$ & $ 10.41-10.98$\\
\hline
\end{tabular}
\begin{tablenotes}
\item[a] Velocity-integrated CO fluxes within the 1\farcs5 aperture. The uncertainties are estimated from the standard deviation of 200 random aperture photometry measurements.
\item[b] Stellar mass and CO-based gas mass within the 1\farcs5 aperture. 
\item[c] The 5$\sigma$ limit of 3 mm continuum-based gas mass within the 1\farcs5 aperture through the gas-to-dust ratio of 120. The range corresponds to the dust emissivity of $\kappa_{850}$=0.4-1.5 g$^{-1}$cm$^2$.
\end{tablenotes}
\end{threeparttable}
\end{center}
\end{table*}

\section{Results and analysis}

We robustly detect CO $J=3-2$ emission in both galaxies as seen in the spatially-averaged spectra within a 1\farcs5~aperture (Figure \ref{fig;spectra}).
We measure total fluxes within a 1\farcs5~aperture in the velocity-integrated maps to derive the CO line luminosities (Table \ref{tab;mass}). 
Both galaxies show a spatial offset between the blue- and red-shifted CO components with a velocity width of 150 km s$^{-1}$ (Figure \ref{fig;spectra}).
The two central positions determine the kinematic major axis of the molecular gas disks.
We also derive line-of-sight velocities by fitting a Gaussian function to the CO line spectrum in each spatial pixel.
The velocity field maps show a monotonic gradient along the kinematic major axis (Figure\ref{fig;spectra}), suggesting rotation of the molecular gas.
In this section, we construct the dynamical model of the dusty star-forming cores through the following three steps: (1) determining a minor-to-major axis ratio ($q=b/a$) of the 870 $\mu$m continuum emission, (2) measuring an effective radius ($R_\mathrm{e}$) of the CO line emission, and (3) exploring the best-fit dynamical model.

\subsection{Spatial extent of dust}
\label{sec;size}

\begin{figure}
\begin{center}
\includegraphics[scale=1]{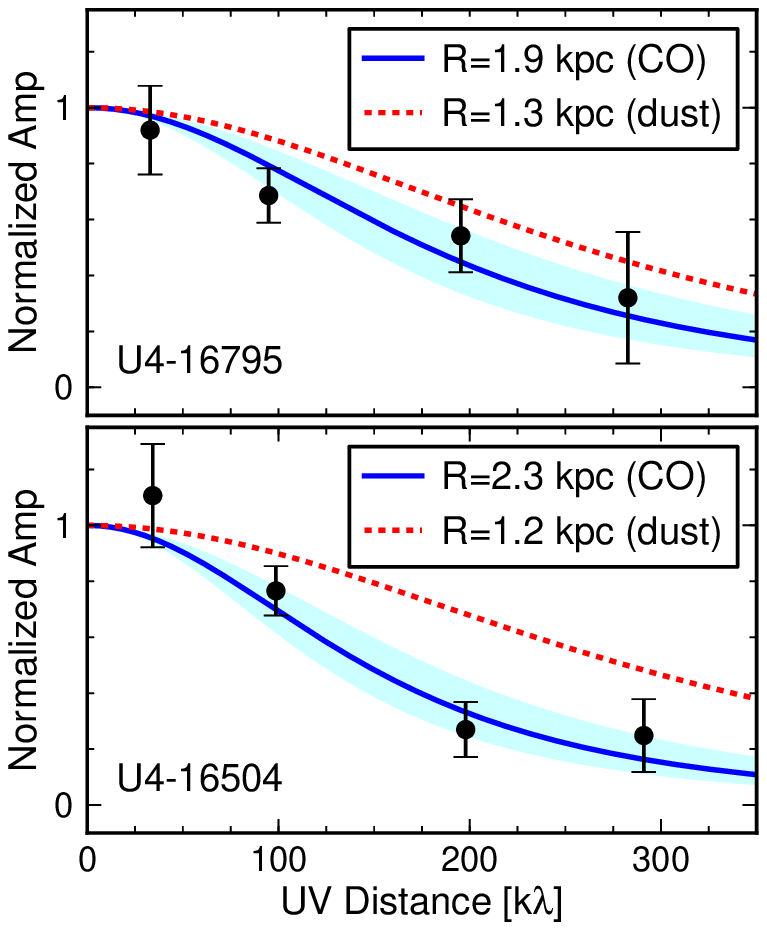}
\caption{Visibility amplitudes of the velocity-integrated CO emission for U4-16795 (top) and U4-16504 (bottom). The blue solid line and the shaded region indicate the best-fitting model and the 1$\sigma$ error, respectively. The red dashed line presents the best-fitting model of the 870 $\mu$m dust emission. The x-axis gives the circularized uv distance.}
\label{fig;visibility}
\end{center}
\end{figure}

0\farcs2-resolution 870 $\mu$m continuum maps are available for both galaxies \citep{2017ApJ...834..135T}.
In SMGs at high-redshift, the dust emission is well described by an elliptical exponential disk \citep{2016ApJ...833..103H}.
We derive effective radii, $R_\mathrm{e}$, along the major axis of the 870 $\mu$m continuum emission assuming an inclined disk with an exponential profile while \citet{2017ApJ...834..135T} have adopted circular disk models ($q=1$).
We use the {\tt UVMULTIFIT} tool to fit the visibility amplitudes to models in the $u-v$ plane \citep{2014A&A...563A.136M}.
For U4-16795, the best-fit values and fitting errors are $R_\mathrm{e}=1.3\pm0.1$ kpc, $q=0.49\pm0.07$ and position angle of PA$_\mathrm{mor}$=35$^\circ \pm5^\circ$.
Note that the morphological major axis of the 870 $\mu$m continuum emission is well aligned with the kinematics major axis of the CO line emission ($|$PA$_\mathrm{mor}$-PA$_\mathrm{kin}|=8^\circ$), supporting ordered rotation \citep{2015ApJ...799..209W}.
We also perform the visibility fitting for U4-16504 with an elliptical disk, but could not obtain meaningful constraints on the axis ratio.
Circular disk models give an effective radius of $R_\mathrm{e}=1.2\pm0.1$ kpc.

\subsection{Spatial extent of molecular gas}
\label{sec;cosize}

Next, we derive effective radii of the CO line emission using the axis ratio of the dust emission.
We fix to $q=0.49$ and PA$_\mathrm{mor}=35^\circ$ for U4-16795, and $q=1$ for U4-16504.
Figure \ref{fig;visibility} shows the observed visibility amplitudes and the best-fit models.
The effective radii are $R_\mathrm{e}=1.9\pm0.4$ kpc for U4-16795 and $R_\mathrm{e}=2.3\pm0.4$ kpc for U4-16504, which are larger than those of the 870 $\mu$m continuum emission.
This result naively suggests that the dust is more concentrated than the molecular gas, which is consistent with negative radial gradients in dust-to-gas mass ratio seen in nearby star-forming galaxies \citep{2011A&A...535A..13M}.
On the other hand, galaxy centers tend to have higher dust temperatures compared to the outer region \citep{2012MNRAS.425..763G}, making the dust mass size larger than the 870 $\mu$m size.
High-resolution ALMA observations at high-frequency band (e.g., 450 $\mu$m) will allow us to determine the radial gradient in dust temperature (section \ref{sec;dust}) and identify if the apparent size difference originates in the dust properties or the intrinsic gradients of dust-to-gas mass ratio.

\subsection{Molecular gas kinematics}
\label{sec;kinematics}

\begin{figure}
\begin{center}
\includegraphics[scale=1]{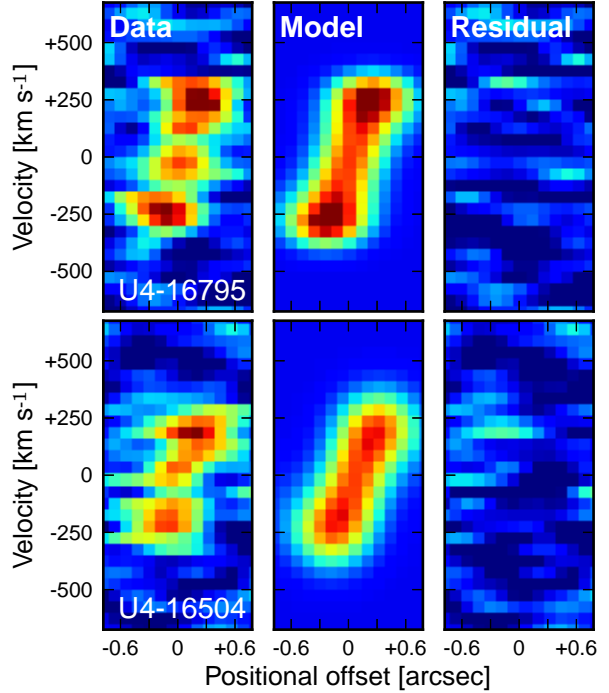}
\caption{Observed position-velocity diagrams of the CO spectra (left). The middle and right panels show the best-fit dynamical model and the residuals between the data and the model, respectively.}
\label{fig;residual}
\end{center}
\end{figure}

We find that both massive galaxies observed here exhibit signatures of disk-like rotation in their CO velocity fields (Figure \ref{fig;spectra}).
Assuming that the molecular gas is in rotating disks, we investigate the kinematic properties by fitting dynamical models to the data in the position-velocity (PV) diagram along the kinematic major axis (Figure \ref{fig;residual}).
We use the {\tt DYSMAL} code \citep{2011ApJ...741...69D} to generate PV diagrams for an exponential disk, spatially convolved with a 0\farcs66$\times$0\farcs55 Gaussian beam.
We take into account the effect of pressure support, reducing the observed rotation velocity \citep[e.g.,][]{2016ApJ...831..149W, 2010ApJ...725.2324B}.
The effective radii of the gas disks are fixed to those measured in Section \ref{sec;cosize}.
For U4-16795, we infer an inclination, $i$, from the axis ratio of the 870 $\mu$m dust continuum emission as $\sin^2 i=(1-q^2)/(1-\mathrm{thickness}^2)$ for symmetric oblate disks with an intrinsic thickness of 0.25 \citep{2014ApJ...792L...6V}. 
The inclination is $\log(1/\sin^2 i)=1.11\pm0.04$ dex (corresponding to $i=64$\degr). 
This uncertainty propagates to the dynamical mass estimate as $M_\mathrm{dyn}\propto1/\sin^2 i$.
As the effective radius and the inclination are fixed, the remaining free parameters in the model are dynamical mass, $M_\mathrm{dyn}$, and local velocity dispersion, $\sigma_0$.
For U4-16504, we adopt the average of possible inclinations ($\langle \sin i\rangle=0.79$) in the case of isotropically oriented disks (see Appendix of \citealt{2009ApJ...697.2057L}). 
As the standard deviation is derived as ($\int(\sin i-\langle \sin i\rangle)^2\sin i~di$/$\int \sin i~di)^{1/2}$=0.22, we adopt the uncertainty of $\Delta \log(1/\sin^2 i)=\pm$0.09 dex.

In the PV diagrams, we use the pixels with a flux above 3$\sigma$ to calculate the chi-squared values between the data and the models.
Figure \ref{fig;residual} shows the best-fit models to minimize the chi-squared value along with the residual maps after subtracting the model from the data.
The best fit models for the two galaxies have dynamical masses  $\log(M_\mathrm{dyn}/M_\odot)=11.10^{+0.07}_{-0.06}$ for U4-16795 and $\log(M_\mathrm{dyn}/M_\odot)=11.19^{+0.12}_{-0.09}$ for U4-16504.
These uncertainties are based on the reduced-chi squared values corresponding to a p-value above 5\%. 
In the dynamical mass measurements, after taking into account the uncertainties of effective radius and inclination, our final uncertainties are +0.13 dex (-0.12 dex) for U4-16795 and +0.17 dex (-0.15 dex) for U4-16504.

We also derive the rotation velocity and local velocity dispersion from the best fit models. We find $v_\mathrm{max}=386^{+36}_{-32}$ km s$^{-1}$ and $\sigma_0=55^{+19}_{-22}$ for U4-16795 and $v_\mathrm{max}=391^{+54}_{-41}$ km s$^{-1}$ and $\sigma_0=96^{+35}_{-31}$ km s$^{-1}$ for U4-16504. 
We note that they have a larger local velocity dispersion than the mean value ($\sigma_0$=50 km s$^{-1}$) in a large sample of rotation-dominated galaxies at $z\sim2$ \citep{2015ApJ...799..209W}.
This means that both dusty star-forming cores are rotation-dominated with 
the ratio of the maximum rotation velocity to local velocity dispersion of $v_\mathrm{max}/\sigma_0=7.0^{+2.5}_{-2.8}$ for U4-16795 and $4.1^{+1.7}_{-1.5}$ for U4-16504.

\section{Dust and gas mass estimates}
\subsection{Dust mass}
\label{sec;dust}

Rest-frame 850 $\mu$m continuum emission is a good indicator of dust mass, $M_\mathrm{dust}$.
For galaxies at $z\sim2$, rest-frame 850 $\mu$m (3 mm in the observed frame) fluxes are often extrapolated from $\sim$1 mm fluxes using a modified-blackbody radiation (MBB) model with dust temperature of $T_d=$25 K and the dust emissivity, $\kappa\propto\nu^{\beta}$, with an index of $\beta=1.8$ \citep[e.g.,][]{2016ApJ...820...83S} as 

\begin{equation}
S_{\nu}=\frac{M_\mathrm{dust} \kappa_{\nu_\mathrm{rest}} B_{\nu_\mathrm{rest}} (T_\mathrm{d})(1+z)}{d_\mathrm{L}^2},
\end{equation}

\noindent
where $d_\mathrm{L}$ is the luminosity distance.
Making 870 $\mu$m maps with the same synthesized beam as the 3 mm maps, we measure peak fluxes of $S_{870\mu\mathrm{m}}=3.5\pm0.1$ mJy beam$^{-1}$ for U4-16795 and $S_{870\mu\mathrm{m}}=2.2\pm0.1$ mJy beam$^{-1}$ for U4-16504.
They correspond to 77\% and 78\% of the total flux within a 1\farcs5 aperture.
Although the MBB models give the extrapolated 3 mm fluxes of $S_{3\mathrm{mm}}=90\pm3~\mu$Jy beam$^{-1}$ for U4-16795 and $S_{3\mathrm{mm}}=55\pm3~\mu$Jy beam$^{-1}$ for U4-16504,
we do not detect the 3 mm continuum emission above 5$\sigma$ significance.

\begin{figure}
\begin{center}
\includegraphics[scale=1]{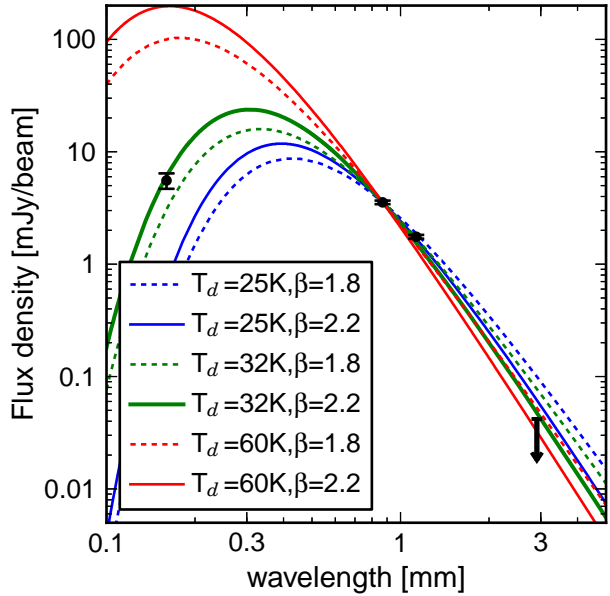}
\caption{Dust continuum SED for U4-16795 with several types of modified blackbody radiation models. The black points are the fluxes measured at PACS 160 $\mu$m, ALMA 870 $\mu$m, and 1.1 mm. The upper limit is given by the 5$\sigma$ flux at ALMA 3 mm.}
\label{fig;dustsed}
\end{center}
\end{figure}

Focusing on U4-16795, we evaluate the assumption of $T_d$ and $\beta$.
We also use the ALMA 1.1 mm flux ($S_{1.1\mathrm{mm}}=1.75\pm0.07~$mJy beam$^{-1}$; \citealt{2015ApJ...811L...3T}). 
Figure \ref{fig;dustsed} shows several MBB models with different dust temperature and emissivity index.
The 5$\sigma$ upper limit at 3 mm rejects the models with $T_d=$25 K, suggesting a higher $T_d$ or a steeper $\beta$.
MBB models with an emissivity index in the usual range ($\beta=1.5-2.0$; \citealt[e.g.,][]{2001MNRAS.327..697D}) require a much higher dust temperature of $T_d>60$ K to explain the faint 3 mm flux 
while they are inconsistent with the 160 $\mu$m flux.
Therefore, we reasonably assume $T_d=$32 K and $\beta=2.2$ to explain all data points. 
In the range of the dust emissivity of $\kappa_{850}$=0.4-1.5 g$^{-1}$cm$^2$ \citep{2003Natur.424..285D}, 
the 3 mm-based dust mass is $\log(M_\mathrm{dust,5\sigma}/M_\odot)=8.30-8.88$ for U4-16795 and $\log(M_\mathrm{dust,5\sigma}/M_\odot)=8.33-8.90$ for U4-16504.
If we use the 870 $\mu$m flux assuming $T_d=$25 K and $\beta=1.8$, the estimated dust mass would become larger by 0.5 dex.

\subsection{Gas mass}
\label{sec;gasmass}

We estimate the CO-based gas masses using two conversion factors, $\alpha_\mathrm{CO}=4.36~M_\odot$ (K km s$^{-1}$pc$^{-2}$)$^{-1}$ (Galactic value) and $\alpha_\mathrm{CO}=0.8~M_\odot$ (K km s$^{-1}$pc$^{-2}$)$^{-1}$ (starburst value; \citealt{1998ApJ...507..615D}). We then compare the gas masses derived
with these two different conversion factors to the dynamical and stellar masses.
The CO $J=3-2$ emission line is assumed to be thermalized \citep{2015ApJ...809..175B}.
These mass measurements are summarized in Table \ref{tab;mass}.
If we adopt the Galactic conversion factor, the baryonic mass fraction, defined as $f_\mathrm{bar}=(M_*+M_\mathrm{gas})/M_\mathrm{dyn}$, exceeds unity.
Monte Carlo simulations incorporating the uncertainties of gas, stellar and dynamical mass show that the probability of the baryonic mass fraction being less than one is 2.3\% for U4-16795 and 17.1\% for U4-16504.
In the case of the starburst conversion factor, the probability is increased to 24.7\% and 55.3\%, respectively.

We also independently constrain the gas mass from the 3 mm-based dust mass, assuming a gas-to-dust ratio of 120 \citep{2008ApJS..178..189W}. 
The 5$\sigma$ upper limit is $\log(M_\mathrm{gas,5\sigma}/M_\odot)=11$ (Table \ref{tab;mass}).
The two independent approaches suggest that the starburst conversion factor is appropriate in the compact dusty star-forming region.
Adopting the starburst conversion factor to derive the CO-based gas mass, we find the gas-to-dynamical mass fraction to be $15^{+4}_{-5}\%$ for U4-16795 and $12^{+4}_{-5}\%$ for U4-16504 and the gas depletion timescale to be $M_\mathrm{gas}/SFR=46\pm11$ Myr and $79\pm46$ Myr, respectively.
When the uncertainties of the conversion factor are taken into consideration, those in the gas mass estimate become larger.

\section{Discussion and Summary}
\label{sec;summary}
Using ALMA observations of CO $J=3-2$ emission, we find that the compact molecular gas in the 
star-forming cores of two massive galaxies is rapidly rotating. 
This has implications for both the formation and subsequent evolution of the cores of massive galaxies. 
The formation mechanism appears to be 
dissipative and the observed rotation indicates that at least some angular momentum is 
preserved in the star-forming molecular gas. Simulations show this could happen due 
to a gas-rich merger or disk instabilities \citep[e.g.,][]{2015MNRAS.450.2327Z,2015MNRAS.449..361W}.

The two massive star-forming galaxies are both rotation-dominated with $v/\sigma=7.0^{+2.5}_{-2.8}$ and $v/\sigma=4.1^{+1.7}_{-1.5}$.
Kinematic studies of massive quiescent galaxies suggest they are rotating \citep{2015ApJ...813L...7N}, although with lower $v/\sigma$ than found here for their progenitors.
Additionally, the descendants of these galaxies in the local universe are slow rotators with $v/\sigma$ nearly an order of magnitude lower \citep{2016ARA&A..54..597C}, 
suggesting that the galaxies we observe need to lose significant net angular 
momentum in the intervening time. 
Our results support a picture in which net angular momentum is initially reduced during 
the quenching process and further during a growth phase by dry mergers. 

We find that the molecular gas of the two massive galaxies at $z=2.5$ is compact with $R_\mathrm{e}\sim2$ kpc.
Such a concentration of star-forming gas is consistent with a scenario in which a wet compaction events (radial transport of gas) could build the cores of massive galaxies \citep{2015MNRAS.450.2327Z}.
The two galaxies host even more compact starbursts with a high SFR surface density as traced by the dust continuum emission, rapidly building up dense cores and transforming the galaxy morphology from disk-dominated to bulge-dominated \citep{2017ApJ...834..135T}.
If the gas mass in these galaxies is as low as our data suggest, we may be witnessing the end of the growth of these dense cores due to star formation.

Our independent measurements of dynamical, stellar, gas and dust mass suggest that the starburst CO-to-H$_2$ conversion factor is appropriate for the spatially resolved cores.
These dense cores are likely to be formed in extreme environments like central regions of local ULIRGs.
The same conclusion is obtained by recent \ci observations of a star-forming galaxy in the similar mass and redshift range \citep{2017arXiv170305764P} and CO observations of starburst galaxies above the main-sequence \citep{2014ApJ...793...19S}.
On the other hand, it is not clear yet that the starburst conversion factor is appropriate for entire galaxies.
\cite{2017arXiv170201140T} investigated the variations in molecular gas properties for 650 galaxies over $0<z<3$ by compiling CO and dust continuum data and does not find the presence of such a large change in the conversion factor. 
A statistical study of the kinematics of the molecular gas is an essential next step for getting a consensus on this issue.
However, high-spatial resolution CO observations suffer from missing flux and are more time-consuming compared to dust continuum observations.
A combination of medium-resolution CO and high-resolution dust continuum observations is reasonable in terms of observing time and a powerful tool for characterizing the dynamical state of molecular gas in distant galaxies.

\

We thank the referee for constructive comments.
This paper makes use of the following ALMA data: ADS/JAO.ALMA\#2013.1.00742.S. ALMA is a partnership of ESO (representing its member states), NSF (USA) and NINS (Japan), together with NRC (Canada), NSC and ASIAA (Taiwan), and KASI (Republic of Korea), in cooperation with the Republic of Chile. 
The Joint ALMA Observatory is operated by ESO, AUI/NRAO and NAOJ. 
K.T. was supported by the ALMA Japan Research Grant of NAOJ Chile Observatory, NAOJ-ALMA-57.
This paper is produced as a part of our collaborations through the joint project supported by JSPS and DAAD under the Japan - German Research Cooperative Program.

\end{document}